\documentclass[aps,prl,twocolumn,superscriptaddress,showpacs,floatfix]{revtex4}
\usepackage{graphicx}
\usepackage{epsfig}
\usepackage{amsmath}
\usepackage{amssymb}

\begin{document}

\title{Structure determination of disordered materials from diffraction data}

\author{Matthew J. Cliffe}
\affiliation{Department of Earth Sciences, Cambridge University,
Downing Street, Cambridge CB2 3EQ, U.K.}

\author{Martin T. Dove}
\affiliation{Department of Earth Sciences, Cambridge University,
Downing Street, Cambridge CB2 3EQ, U.K.}

\author{D. A. Drabold}
\affiliation{Department of Physics and Astronomy, Ohio University, Athens, Ohio 45701-2979}

\author{Andrew L. Goodwin}
\email{andrew.goodwin@chem.ox.ac.uk}
\affiliation{Department of Earth Sciences, Cambridge University,
Downing Street, Cambridge CB2 3EQ, U.K.}
\affiliation{Department of Chemistry, Inorganic Chemistry Laboratory, University of Oxford, South Parks Road, Oxford OX1 3QR, U.K.}

\date{\today}
\begin{abstract}
We show that the information gained in spectroscopic experiments regarding the number and distribution of atomic environments can be used as a valuable constraint in the refinement of the atomic-scale structures of nanostructured or amorphous materials from pair distribution function (PDF) data. We illustrate the effectiveness of this approach for three paradigmatic disordered systems: molecular C$_{60}$, a-Si, and a-SiO$_2$. Much improved atomistic models are attained in each case without any \emph{a-priori} assumptions regarding coordination number or local geometry. We propose that this approach may form the basis for a generalised methodology for structure ``solution'' from PDF data applicable to network, nanostructured and molecular systems alike.
\end{abstract}


\maketitle

Many materials of fundamental importance possess structures that do not exhibit long-range periodicity: examples include metallic and covalent glasses \cite{Byrne_2008}, amorphous biominerals \cite{Weiner_2005}, the so-called ``phase-change'' chalcogenides of DVD-RAM technology \cite{Sun_2006}, and amorphous semiconductors such as a-Si and a-Ge \cite{Armatas_2006}. The absence of Bragg reflections in the diffraction patterns of these materials precludes the use of traditional crystallographic techniques as a means of determining their atomic-scale structures. Yet it is clear that these materials do possess well-defined local structure on the nanometre scale \cite{Billinge_2007}; moreover it is often this local structure that is implicated in the particular physical properties of interest \cite{Billinge_2004}. For this reason, the development of systematic information-based methodologies for the determination of local structure in disordered materials remains one of the key challenges in modern structural science \cite{Juhas_2006}.

Historically, local structure has been studied experimentally using two principal approaches: (i) the diffraction techniques of neutron and x-ray total scattering, from which the distribution of interatomic separations can be measured via the pair distribution function (PDF) \cite{Egami_2003}, and (ii) resonance and spectroscopic methods (NMR, EXAFS, IR, Raman) that yield information concerning the number and population of distinct atomic environments, together with (in favourable cases) metal-coordination/molecular geometries \cite{Brodsky_1978,Mullerwarmuth_1982}. These techniques afford a rich body of information, and over the past 5--10 years a number of sophisticated methods of analysis have emerged that aim to derive structural models via fitting to these experimental data. The Reverse Monte Carlo (RMC) \cite{McGreevy_2001} and Empirical Potential Structure Refinement (EPSR) \cite{Soper_2005} methods have been used widely in the glass and amorphous materials community, while the PDFfit \cite{Proffen_1999} and ``Liga'' \cite{Juhas_2008} methods have been applied more recently to nanostructured solids such as C$_{60}$ \cite{Juhas_2006} and ferrihydrite \cite{Michel_2007}---systems that present similar crystallographic challenges.

There is, however, a fundamental problem: markedly different structural models can be equally consistent with the same PDF data \cite{Gereben_1994}. Moreover, the task of fitting simultaneously to PDF and spectroscopic data is almost always either too computationally demanding or in fact not quantitatively possible. Taken together, these factors have meant that it is often difficult to determine the atomic-level structure of these materials, and that there is no ``routine'' information-based approach analogous to those for crystalline materials.

In this Letter, we show that this problem can largely be solved by using information gained via spectroscopy---the number and population of distinct atomic environments---to guide refinement of experimental PDF data. Structural refinement based on reproducing the experimental PDF alone is, in general, not sufficiently well-constrained to produce models that reflect the ``true'' local structure in a material; however, if refinement is forced also to reflect the correct number and distribution of atom environments then convergence on the correct local structure usually follows. This approach is easily implemented and generic. Moreover, we show that successful refinement can be initiated using entirely random atomistic models and, in being driven wholly by experimental data, one avoids any other \emph{a-priori} assumptions concerning \emph{e.g.}\ coordination numbers or geometries. While our focus lies on proof-of-principle at this stage, our results show that routine information-based structure determination of disordered materials is now a viable prospect.

Our paper is arranged as follows. We begin by describing the particular implementation of our methodology through a ``variance''-based term in the cost function used to drive PDF refinement. Three principal case studies follow: nanoparticulate C$_{60}$ (single cluster; one atom environment), amorphous silicon (continuous network solid; one atom environment) and amorphous silica (continuous network solid; two atom environments). In all three instances we show that a conventional RMC approach fails to obtain the correct structure solution---often spectacularly---but that inclusion of the variance term is sufficient to recover almost-perfect models of material structure in each case. We conclude by discussing a number of different possible implementations of our underlying methodology.

In outlining our methodology, it is useful to consider first the simplest type of disordered material: namely a phase that contains a single atom type and for which spectroscopy indicates a single atom environment. The existence of a single atomic environment demands that structural correlation functions calculated for different individual atoms within the material should take similar forms. In order to recast this statement with specific reference to the PDF, we first define atomic PDFs $p_j(r)$ for an atomistic model such that the ``bulk'' (measurable) PDF $G(r)$ corresponds to the average $\langle p(r)\rangle$ taken over all atoms $j$. Then the existence of a single atom environment dictates a similarity $p_j(r)\sim p_{j^\prime}(r)\sim G(r)$ for all atoms $j,j^\prime$. Whereas a standard PDF-based structure refinement would involve minimising a function of the form

\begin{equation}\label{oldchi2}
\chi^2=\sum_r[\langle p(r)\rangle-G_{\rm expt}(r)]^2,
\end{equation}

\noindent what we would propose is the alternative cost function 

\begin{equation}\label{newchi2}
\chi^2=\frac{1}{N}\sum_j\sum_r[p_j(r)-G_{\rm expt}(r)]^2.
\end{equation}

\noindent Note that in this reformulation one obtains $\chi^2=0$ if and only if the model PDF matches $G_{\rm expt}(r)$ \emph{and} each individual $p_j(r)$ has the same form. It is straightforward to show that the new penalty function $\chi^2$ of \eqref{newchi2} is in fact equal to that of \eqref{oldchi2} plus a variance term $\chi^2_{\rm Var}=\langle p(r)^2\rangle-\langle p(r)\rangle^2$. What the spectroscopic result suggests is to add to a conventional PDF refinement a term that penalises variance in local coordination environments; for this reason we are terming our approach an INVariant Environment Refinement Technique (INVERT).

In practice, the individual $p_j(r)$ for a static atomistic model consist of a series of delta functions, and in order to obtain a well-behaved variance term, it is necessary to adopt a modified formulation such as:

\begin{equation}\label{varchi2}
\chi^2_{\textrm{Var}}=\frac{1}{N}\sum_j\sum_i\frac{\left[d_j(i)-\langle d(i)\rangle\right]^2}{\langle d(i)\rangle^2},
\end{equation}

\noindent where $d_j(i)$ measures the distance from atom $j$ to its $i$-th neighbour, and $\langle d(i)\rangle$ is the average such distance over all atoms $j$. The term in the denominator of Eq.~\eqref{varchi2} appears in order to account for the fact that the number of neighbours at a distance $d$ scales with $d^2$ \cite{othermodifications}. The extension to multiple atom types and/or atom environments is straightforward. A separate variance term is included for each different pair of atom types (A and B, say); the form of each individual term is the same as in Eq.~\eqref{varchi2} except that the $d_j(i)$ will refer to $i$-th neighbour of type B around the $j$-th atom of type A, and so on.

At this point we emphasise that no assumption has been made regarding the actual distribution of neighbours around each atom---only that this distribution should be similar for equivalent atoms. Moreover, we are able to constrain the partial PDF functions for multi-component systems despite the experimental PDF data representing a sum over these separate contributions.

\begin{figure}
\begin{center}
\includegraphics{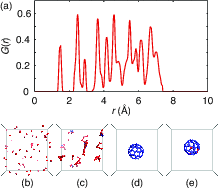}
\end{center}
\caption{\label{fig1}RMC refinement of the experimental PDF of C$_{60}$: (a) the neutron PDF itself, corrected to remove inter-molecular correlations as described in Ref.\ \onlinecite{Juhas_2008}; (b) a random starting configuration of 60 carbon atoms; (c) a typical configuration produced by conventional RMC refinement of either idealised or experimental PDF data; and those produced by INVERT+RMC using (d) idealised, and (e) experimental PDF data. In panels (b)--(e), atoms with three nearest neighbours are coloured blue and others are coloured red.}
\end{figure}

We have chosen C$_{60}$ as a simple first case study, not least because the task of determining its well-known icosahedral structure from the experimental PDF [Fig.~\ref{fig1}(a)] has recently been highlighted as a benchmark challenge in nanostructure determination \cite{Juhas_2008}. As straightforward as the task might seem, conventional RMC refinement from a random starting configuration [Fig.~\ref{fig1}(b)] fails entirely, giving a set of small clusters that contains only a few of the real set of interatomic separations [Fig.~\ref{fig1}(c)]. The same result is obtained even if idealised PDF data are used.

The INVERT modification exploits the experimental NMR result that C$_{60}$ contains a single C environment \cite{Johnson_1990}. Clearly the RMC configuration in Fig.~\ref{fig1}(c) violates this property, and so would now give rise to a large $\chi^2_{\rm Var}$ term that will help drive refinement forward. Indeed, INVERT+RMC refinement from the same random starting configuration gives the correct solution for idealised data [Fig.~\ref{fig1}(d)] and a near-perfect solution for the experimental neutron PDF data of Ref.~\onlinecite{Juhas_2008} [Fig.~\ref{fig1}(e)]. We note that such a result has only ever been achieved previously using the highly-sophisticated cluster optimisation methods of the ``Liga'' algorithm or using genetic algorithms based on the principle of mating or crossover (the latter only giving correct solutions in 56\,\% of attempts) \cite{Juhas_2006,Hartke_1999,Deaven_1995}. Here, INVERT+RMC consistently obtains a topologically-identical solution from random starting coordinates in approximately 2\,000--4\,000 accepted moves.

We find the extension to a cluster with two atom environments---namely, S$_{12}$ \cite{Steidel_1981}---enjoys similar success \cite{s12note}. Videos that illustrate the refinement process for C$_{60}$ and S$_{12}$ are provided as supporting information \cite{Movies}.

The paradigmatic ``stumbling block'' for RMC, however, has always been amorphous Si, whose structure is believed to consist of a continuous random network (CRN) of tetrahedral Si centres \cite{Drabold_2009}. Rather than generating a network of four-fold-coordinated Si atoms, RMC refinements of a-Si PDF data yield configurations with unphysically-broad distributions of Si coordination numbers \cite{Gereben_1994}. This is allowed because, crudely speaking, a pair of atoms of coordination numbers three and five will contribute to the average PDF indistinguishably from two four-fold coordinated atoms, and yet the former state is statistically more likely during a sequence of random moves. Various work-arounds have been proposed and implemented (\emph{e.g.}\ constraining coordination numbers to equal four), and in the most favourable of cases these yield CRNs comparable to those obtained from bond-switching (Wooten-Weaire-Winer, ``WWW'' \cite{Wooten_1985}) methods, molecular dynamics and \emph{ab-initio} calculations \cite{Drabold_2009}.

\begin{figure}
\begin{center}
\includegraphics{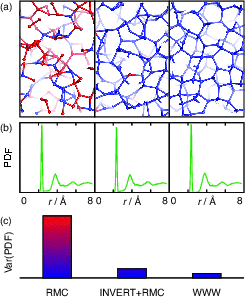}
\end{center}
\caption{\label{fig2}Comparison of a-Si configurations obtained using (left) ``Native RMC'' and (centre) ``INVERT+RMC''' refinement with (right) the trusted ``WWW''model of Ref.~\onlinecite{Wooten_1985}. (a) Slices of the configurations themselves, with four-coordinate Si coloured blue and others coloured red. (b) The PDFs calculated from each configuration, which are essentially identical. (c) Corresponding PDF variances calculated using Eq.~\eqref{varchi2}.}
\end{figure}

However, there is a sense in which one recovers from these approaches only the very information already used to generate the constraints: if the coordination number is constrained to equal four during refinement, then four-fold coordination cannot be considered an independent result. Consequently, our motivation for considering a-Si as a second case study was primarily to determine whether, by using the evidence for a single Si environment from NMR studies \cite{Shao_1990}, INVERT+RMC refinement could yield reasonable structural models without recourse to explicit coordination number constraints.

First, a conventional RMC refinement was performed using $G(r)$ ``data'' generated from the trusted WWW model of Ref.~\onlinecite{Wooten_1985}. The starting configuration was a random collection of 512 Si atoms in a cubic box of side 21.7\,\AA. Refinement gave a highly-disordered configuration that displayed all the hallmarks of previously-described problematic RMC studies \cite{Gereben_1994}: only 27\,\% of Si atoms are four-fold coordinated, there are substantial density variations, and large numbers of unphysical Si$_3$ ``triangles'' [left-hand panel of Fig.~\ref{fig2}(a)]. In contrast, a parallel INVERT+RMC refinement achieved an almost perfect coordination distribution (95\,\% four-fold). The improvement extended even to the higher-order correlations (discussed in more detail below): in particular, the number of Si$_3$ triangles is halved, and the density distribution is much more even. Inspection of the configuration itself [centre panel of Fig.~\ref{fig2}(a)] now reveals obvious similarities to the trusted WWW model [right-hand panel of Fig.~\ref{fig2}(a)]. The PDF itself is relatively insensitive to this fundamental improvement in local structure modelling [Fig.~\ref{fig2}(b)], while the variance term of Eq.~\eqref{varchi2} clearly acts a much better figure-of-merit [Fig.~\ref{fig2}(c)].

\begin{figure}
\begin{center}
\includegraphics[width=4cm]{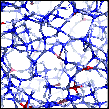}
\end{center}
\caption{\label{fig3}A slice of the information-driven CRN model of a-SiO$_2$ obtained using INVERT+RMC refinement as described in the text. Si and O atoms shown in dark and light colours, respectively; atoms in shades of blue have the expected coordination numbers of 4 (Si) and 2 (O), while the few in shades of red have incorrect coordination numbers.}
\end{figure}

Similar results are obtained for amorphous SiO$_2$, which is a conceptual extension in that it contains two distinct atom environments: that of the Si atoms and that of the O atoms. Experimental neutron PDF data were taken from Ref.~\onlinecite{Tucker_2005}, and starting configurations generated from a random distribution of 64 Si atoms and 128 O atoms in a periodic cubic box of side 14.37\,\AA. RMC refinement both with and without the INVERT modification gave excellent fits to the PDF, but the INVERT+RMC model had a much higher percentage of fourfold Si coordination (97\% \emph{vs}.\ 59\% for the RMC-only configuration). Indeed, we believe the INVERT+RMC configuration to be the first information-based CRN model of a-SiO$_2$ [Fig.~\ref{fig3}].

The INVERT methodology is by no means applicable only to RMC refinement. Our focus on RMC in this Letter arises from a desire to demonstrate the effectiveness of the INVERT approach using a refinement method that is known to favour \emph{disorder}. The incorporation of variance-based cost functions in any refinement approach is straightforward, and such a modification to more sophisticated PDF fitting approaches than RMC, \emph{e.g.}\ as suggested in Ref.~\onlinecite{Juhas_2008}, might reasonably be expected to produce even more realistic configurations.

Speaking more generally, we would note that the concept of local invariance encompasses more than minimising the PDF variance alone. One can imagine, for example, that minimising the variance in higher-order correlation functions, such as angle distributions, coordination geometry, and CRN ring statistics might also improve refinement further. Importantly, these constraints can be implemented despite the functions not being measurable experimentally. In practice, however, we have found that the calculation of higher-order correlation functions is too computationally-demanding for speedy refinement at this stage; the extension to constraining geometric invariance using spherical harmonics and/or the triplet distribution function is an approach we hope to pursue further in the near future.

\subsection*{Acknowledgements}

We gratefully acknowledge financial support from Trinity College, Cambridge to A.L.G., from the EPSRC (UK) to A.L.G.\ and M.J.C., and from the US NSF to D.A.D.\ under grant DMR 09-03225. We thank D.\ A.\ Keen (Rutherford Appleton Laboratory) for useful discussions, and acknowledge the University of Cambridge's CamGrid infrastructure for computational resources.

\end{document}